\documentclass[letterpaper,twoside,twocolumn,american,pra,aps,showpacs,floatfix]{revtex4}
\usepackage[T1]{fontenc}
\usepackage[latin1]{inputenc}
\usepackage{amsmath}
\usepackage{graphicx}
\usepackage{amssymb}

\makeatletter

\usepackage{babel}
\makeatother
\begin{document}

\title{Nuclear spin qubits in a pseudo-spin quantum chain}

\author{E. \surname{Novais}}

\affiliation{Department of Physics, Boston University, 590 Commonwealth Ave.,
Boston, MA, 02215}

\author{A.~H. \surname{Castro~Neto}}

\affiliation{Department of Physics, Boston University, 590 Commonwealth Ave.,
Boston, MA, 02215}

\date{\today{}}


\begin{abstract}
We analyze a quantum computer (QC) design based on nuclear spin
qubits in a quasi-one-dimensional (1D) chain of non-Kramers doublet
atoms. We explore the use of spatial symmetry breaking to obtain control
over the local dynamics of a qubit. We also study the decoherence mechanisms
at the single qubit level and the interactions mediated by the magnetic
media. The design can be realized in $\textrm{PrBr}_{\textrm{3}-x}\textrm{F}_{x}$
with nuclear magnetic
resonance (NMR) techniques. 
\end{abstract}
\maketitle

\section{Introduction}

Nuclear magnetic resonance is the framework of
a very promising quantum computing architecture \cite{NAG97}.
NMR is a natural choice because nuclei are protected from many sources
of decoherence, and therefore produce robust qubits. Successful
realizations of quantum algorithms implemented on a NMR quantum computer
have been realized in liquid solutions of
molecules \cite{LMKV01,LMKV02}.
Nevertheless, a liquid NMR QC is not easily scalable, that is,
there is a practical limit in the number of qubits that can be constructed
in a molecule. From a handful of qubits already achieved one must scale
the QC to several thousands before a non-trivial algorithm
can be run \cite{LMKV01,LMKV02}. Though other limitations can
also be argued to the use of NMR \cite{NLSP01}, scalability is
an undeniable problem.

A possible route to deal with the scalability problem is to
consider NMR in crystals \cite{TDL+00}.
There are several different proposed designs,
but all of them share two common elements.
Firstly, a gradient magnetic field is used to shift the nuclear resonance
frequencies of different nuclei, allowing qubits to be addressed independently.
Secondly, as the number of qubits increases, a second decoherence channel is
introduced by the low energy excitations of the interacting qubits.
In any crystal, the direct dipolar interaction between nuclei produces 
secular broadening.
To a certain extent this broadening can be reduced by NMR techniques.
Thus, it is usually assumed that a \emph{perfect selective decoupling}
of the qubits from the dipolar interaction can be achieved. 

Although very promising, there are technical problems with the use
of NMR in crystals.
For example, in the proposed materials $\textrm{CaF}_{\textrm{2}}$
and $\textrm{MnF}_{\textrm{2}}$, qubits are the nuclear spin $1/2$
of the F ions \cite{TDL+00}. To obtain a measurable frequency shift
from one qubit to another a homogeneous gradient field of more than
$1T/\mu m$ is required. The obvious solution is to separate qubits
from each other. However, by distancing the qubits to work with an experimentally
feasible value of the field gradient, another problem is created by weakening the
qubit-qubit interactions.

Interacting qubits are a necessary condition for quantum computation.
A quantum algorithm is a sequence of unitary transformations
in the Hilbert space spanned by all the qubits.
A given transformation in a subspace of
\emph{n} qubits is called a \emph{n}-qubit gate.
A quantum computing scheme must provide a complete
set of such quantum gates, in other words, it must be possible
to construct any unitary transformation with a sequence of building
block operations provided by the design. 
One of the most useful results
in quantum information theory is that from all \emph{one}-qubit gates
and \emph{almost} any \emph{two}-qubit gate is possible to find
a complete set of gates \cite{SL95}.
In a NMR QC, the \emph{one}-qubit gates are easily produced.
The \emph{two}-qubit gate is the time evolution of two qubits under
an interaction.

The viability of a solid state NMR QC
relies on interactions available to construct the \emph{two}-qubit gate
and the correspondent decoherence times.
On the one hand,
in $\textrm{CaF}_{\textrm{2}}$, the only available interaction is
the direct dipolar coupling between nuclear moments. In most cases
this interaction is effectively short ranged for quantum computational
purposes.
The small nuclear moments and the $1/r^{3}$ dependence makes the
operation time of
a gate (composed by two qubits far apart) much larger than the
decoherence times.
On the other hand, in
$\textrm{MnF}_{\textrm{2}}$,
the relevant interaction is the 
Suhl-Nakamura coupling \cite{HS58}.
This is an indirect coupling of nuclear spins  mediated by magnons of the
$\textrm{Mn}$ electronic spins. Below its N\'{e}el temperature the magnon spectrum has
 a gap. At the same time that a gap reduces decoherence,
it implies that the interaction strength has an exponential decay with the distance.
Thus, it is unlikely that a considerable separation between qubits
can be obtained in both cases.
The search for long range interactions has motivated several recent
publications \cite{DMVP01,AGTS+03}.
Unfortunately, long range interactions
are tied to low energy modes and, consequently, short decoherence times.

In this paper, we discuss nuclear-nuclear interactions mediated
by an anisotropic quantum pseudo-spin chain. We analyze how the breaking
of spatial symmetries in a system of non-Kramers ions can be used
to gain control over local properties of a QC. We show that one
can reduce decoherence and/or construct different \emph{two}-qubit gates
as a function of external electromagnetic fields.
Although our ideas are general, we
propose a specific realization in the compounds
$\textrm{PrCl}_{\textrm{3-x}}\textrm{F}_{\textrm{x}}$
and $\textrm{PrBr}_{\textrm{3-x}}\textrm{F}_{\textrm{x}}$.
Both materials are equally suitable
to our discussion, but we use the parameters of the latter in
our estimates. We start by summarizing the properties of the parent compound,
$\textrm{x=0}$. Subsequently, we discuss the chemical doping
with F. Finally, we explore the use of the nuclear spin from the F
ions as qubits. 

\section{the physics of $\textrm{PrBr}_3$ and the construction of qubits}

\begin{figure}[bth]
\includegraphics[ width=0.9\columnwidth  ]{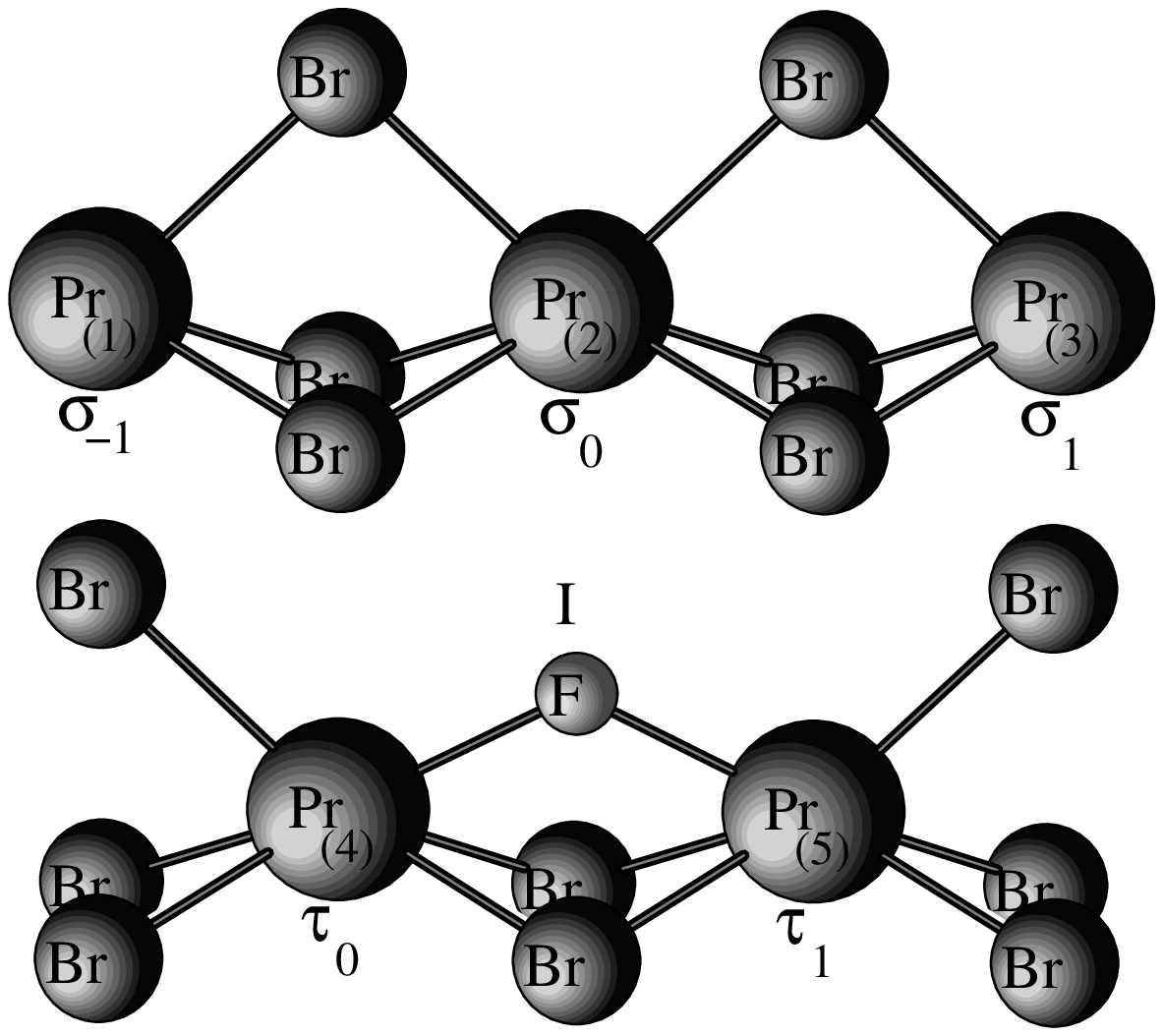}
\caption{\label{cap:PrBr3-xFx}Two adjacent chains in
$\textrm{PrBr}_{3-x}\textrm{F}_{x}$.}
\end{figure}

$\textrm{PrBr}_{\textrm{3}}$ is a 1D ionic insulator made out of
Pr chains separated by $5\textrm{Å}$. The Pr ions are subjected to
a crystal field with $C_{3h}$ symmetry. Their ground state is a 
non-Kramers doublet that is separated from the first excited state by
a gap of $17\textrm{K}$ \cite{BSBH+87}. A Jahn-Teller transition
takes place at $0.1\textrm{K}$ \cite{SSRLA92}, it lifts the doublet 
degeneracy, and sets a low temperature limit to the applicability of this material
to our design.
A convenient way to model this system is via a pseudo-spin
$1/2$ representation \cite{AABB70}. We focus on
the physics of two adjacent chains and we label the pseudo-spins
of each of these chains as $\tau ^{z}$ and $\sigma ^{z}$
(see Fig.~\ref{cap:PrBr3-xFx}).
The single ion Hamiltonian at site $i$ is written as
\begin{eqnarray*}
H_{\textrm{ion}} & = & \sum _{i}\hslash \gamma _{s}^{z}B_{z}S_{i}^{z}+g_{s}^{x}E_{x}S_{i}^{x}+g_{s}^{y}E_{y}S_{i}^{y},
\end{eqnarray*}
where $\overrightarrow{S}=\left\{ \overrightarrow{\sigma },\overrightarrow{\tau }\right\} $,
$\gamma _{s}^{z}=1.4\times 10^{11}\textrm{T}^{\textrm{-1}}\textrm{s}^{\textrm{-1}}$,
$\vec{B}$ is an external magnetic field and $\vec{E}$ an applied
electric field \cite{BSBH+87}. We are unaware of published values
for the electric dipolar constants in $\textrm{PrBr}_{\textrm{3}}$,
however they should not be very different from the ones in $\textrm{PrCl}_{\textrm{3}}$
where $g_{s}^{x,y}=4.0\times 10^{-31}\textrm{Cm}$ \cite{JPH76}. It is important
to stress that there is no off-diagonal matrix element that couples
the doublet state to the magnetic field. Therefore, a magnetic field
cannot induce transitions between the doublet states.
The ionic magnetic moments are coupled
by a dipolar term, however the most relevant contribution to the interaction
Hamiltonian comes from transitions due to the transverse electric
dipoles that are strongly coupled to the lattice. Although the only real
magnetic moment is oriented along the chain ($z$ direction), this family of compounds is regarded
as $\textrm{XY}$ chains described by the Hamiltonian
\begin{eqnarray}
H_{\textrm{xy}} & = & J_{\perp }\sum _{i}S_{i}^{x}S_{i+1}^{x}+S_{i}^{y}S_{i+1}^{y},\label{eq:Hxy}
\end{eqnarray}
where $J_{\perp }\approxeq 3\textrm{K}$ \cite{SSRLA+91}.

In order to construct a qubit, we propose the use of the nuclear spins
of $\textrm{F}$ ions in the diluted salt $\textrm{PrBr}_{\textrm{3-}x}\textrm{F}_{x}$.
There are two main components to nuclear decoherence, connected with
the two strongest interactions that a nucleus is subjected to: the
electric quadrupolar and the magnetic dipolar \cite{AA}.
We are ultimately interested in the decoherence channels in a F
nucleus in  $\textrm{PrBr}_{\textrm{3-x}}\textrm{F}_{\textrm{x}}$.
Nuclear quadrupole
resonance experiments have measured  $T_{1,2}$
for the $\textrm{Br}$ nuclei in the parent compound.
They established fairly well that
the spin-lattice relaxation time, $T_{1}$, is due to magnetic interactions
\cite{MDRLA83,SSRLA+91},
and it is of order of $100 ms$ at 1K.
Moreover, the nuclear spin-spin decoherence time, $T_{2}$, was found to be
$\approxeq 40\mu s$ at 1K. The decoherence sources that lead to this 
value for $T_{2}$ are not yet well understood \cite{SSRLA92}.
If we use the Van Vleck formula \cite{AA} to estimate
the secular broadening of resonance
lines, we find that  the direct dipolar interaction among the nuclei leads to a
broadening of the order of $10^{2}\mu \textrm{s}$.
Further considering the quadrupolar effects
it is clear that the direct
dipolar interaction gives a sizable contribution to decoherence.
Thus, as usual in solid state NMR designs, we can conclude that
decoupling is very important
in order to make this family of compounds useful to a QC.

Each F introduces a local lattice distortion, hence lowering the crystal
field symmetry at neighboring Pr ions.
The distortion introduced by the $\textrm{F}$
ion has its strongest effect on the Pr ions labeled $2$, $4$, $5$
in Fig.~(\ref{cap:PrBr3-xFx}). In the pseudo-spin representation,
a local symmetry breaking corresponds to the addition of 
transverse fields, $\bar{\Delta }$ and $\Delta $, on each one of these
sites. Moreover, the $\textrm{Pr}_{\left(\textrm{4,5}\right)}$ no
longer have a plane of inversion perpendicular to the chain axis. 
Thus, these ions can develop electric dipoles perpendicular to
that plane. The Hamiltonian for the pseudo-spin chains
can be written as
\begin{eqnarray}
H_{\textrm{Pr}} & = & H_{\textrm{ion}}+H_{\textrm{x}y}+\Delta \sigma _{0}^{x}+\bar{\Delta }\left(\tau _{0}^{x}+\tau _{1}^{x}\right)\nonumber \\
 & + & g_{s}^{z}E_{z}\left(\tau _{0}^{z}-\tau _{1}^{z}\right).\label{eq:fullhamispin}
\end{eqnarray}

We consider the case where
$\left(\bar{\Delta },\, \Delta \right)\ll
 \max \left(\hslash \gamma _{s}^{z}B_{z},k_{B}T\right)<J_{\perp }$,
otherwise the moments at $\textrm{Pr}_{\left(2,4,5\right)}$ would be completely
quenched by the symmetry breaking and the analysis below would need
to be extended to include next near neighbor interactions.
Notice that in Eq.~(\ref{eq:fullhamispin}) 
the transverse fields introduce matrix elements between the two magnetic
states of $Pr_{\left(2,4,5\right)}$. Thus, an oscillating magnetic
field parallel to the chain axis would reveal two distinct resonant
lines, $\omega _{\bar{\Delta }}$ and $\omega _{\Delta }$, associated
with the splitting of the $\textrm{Pr}$ doublet state.

\section{the qubit Hamiltonian}

The use of $\textrm{F}$ as a qubit has two advantages.
There is no decoherence due to electric field gradients because 
it does not have a quadrupolar moment.
In addition, there is only one isotope of $\textrm{F}$ in nature,
so all qubits experiencing the same magnetic field are identical.
By assuming perfect decoupling, we can disregard
the direct dipolar interaction between nuclei.
This is a much less stringent condition than in other NMR QC schemes
because the qubit resonance frequency is very distinct from the other ions.
Therefore, straightforward pulse sequences can be used to perform the decoupling.  
The remaining contribution to the
nuclear Hamiltonian comes from the magnetism of the surrounding Pr
atoms. Hence, the nuclear hyperfine interaction
of each F ion in first approximation can be written as
\begin{eqnarray}
H_{\textrm{F}} & = & \left[\hslash \gamma _{N}B_{z}+d\left(\sigma _{0}^{z}-\frac{\tau _{0}^{z}+\tau _{1}^{z}}{2}\right)-\tilde{d}\left(\sigma _{-1}^{z}+\sigma _{1}^{z}\right)\right]I^{z}\nonumber \\
 & + & 3d\left(\tau _{0}^{z}-\tau _{1}^{z}\right)I^{x}+\sqrt{2}\tilde{d}\left(\sigma _{-1}^{z}-\sigma _{1}^{z}\right)I^{y},\label{eq:hf-1}
\end{eqnarray}
where $d=(\mu _{0}\hslash ^{2}\gamma _{s}^{z}\gamma _{N})/(4\pi
r_{0}^{3})\approxeq 10^{-4}\textrm{K}$,
$\tilde{d} \approx d/5$, $\gamma _{N}=25\times 10^{7}\textrm{T}^{\textrm{-1}}\textrm{s}^{\textrm{-1}}$,
and $\vec{I}$ is the nuclear spin-$1/2$ operator of the $\textrm{F}$
nucleus.

The pseudo-spin physics described by Eq.~(\ref{eq:fullhamispin})
presents us with a very interesting situation. An applied magnetic
field with frequency $\omega _{\bar{\Delta }}$ and/or an electric
field, $E^{z}$, only affect the $\textrm{Pr}_{\left(\textrm{4,5}\right)}$,
and therefore can be used to act locally in the qubit. For example,
a sufficiently large electric field forces $\tau _{0}$ and $\tau _{1}$
into a singlet configuration, freezing their dynamics. The net result
is decoupling of the F ion from the $\tau $-chain. In this case the
hyperfine Hamiltonian simplifies to

\begin{eqnarray}
H_{\textrm{F}} & \approxeq  & \left[\hslash \gamma _{N}B_{z}+d\sigma _{0}^{z}-\tilde{d}\left(\sigma _{-1}^{z}+\sigma _{1}^{z}\right)\right]I^{z}\nonumber \\
 & + & \sqrt{2}\tilde{d}\left(\sigma _{-1}^{z}-\sigma _{1}^{z}\right)I^{y}.\label{eq:hype2}
\end{eqnarray}
This is a particularly interesting effect. It cancels the strongest
transverse part of Eq.~(\ref{eq:hf-1}), and consequently, corresponds
to a reduction in the dissipation rates $T_{1,2}^{-1}$.

\subsection{Dissipation rates}

In order to estimate  the dissipation rates due to the Pr magnetic moments,
we will focus on the low energy physics of
Eq.~(\ref{eq:fullhamispin}). Therefore, we can use Abelian bosonization\cite{Ian} to obtain
simple analytical expressions for $T_{1,2}^{-1}$.

Bosonization is a well stablished method to study spin chains.
In a concise way, we first use the Jordan-Wigner transformation,
mapping the pseudo-spins in spinless fermions.
Then, we linearizing the dispersion relations around the two Fermi points,
$p_{F}=\arccos \left(\hslash \gamma_{s}^{z}B_{z}/J_{\perp }\right)$,
and define the 
Fermi velocity $v=J_{\perp }\sin \left(p_{F}\right)$.
The result is that $H_{\textrm{xy}}$ can be re-written
as a free bosonic Hamiltonian.
In this language, it is straighforward to evaluate the
pseudo-spin correlation function at zero temperature\cite{Ian}

\begin{eqnarray}
\left\langle S^{z}_{j}(\tau)S^{z}_{0}(0\right\rangle & = & \frac{1}{2\pi^{2}}
\frac{x^2-(v\tau)^2}{(x^2+(v\tau)^2)^2}\nonumber\\
& & \frac{\cos(2p_{K}x)}{2\pi^{2}}\frac{1}{x^2+(v\tau)^2}\label{correlation},
\end{eqnarray}
where $\tau$ is the imaginary time, $x=a_{0}j$ and $a_0 \approx 4.4\textrm{Å}$
is the lattice spacing.

For a sufficiently large magnetic field ($B_{z}\gg 0.1\textrm{T}$),
$T_{1}^{-1}$ is given by\cite{slichter}
\begin{eqnarray}
T_{1}^{-1} & = & \frac{1}{2}\int_{-\infty}^{\infty}dt^{\prime}
\left\langle H_{\perp}(t) H_{\perp}(t+t^{\prime}) 
\right\rangle e^{-i\omega_{0}t^{\prime}},
\label{slichter}
\end{eqnarray} 
where, if we focus in the regime described by  Eq.~(\ref{eq:hype2}), we defined
\begin{eqnarray*}
\omega_{0} & = & \gamma_{N} B_{z},\\
H_{\perp} & = & \sqrt{2} \tilde{d}\left(\sigma _{-1}^{z}-
\sigma _{1}^{z}\right).
\end{eqnarray*}
An equivalent expression for $T_{2}$ is obtained when we match
the results of a random phase approximation (RPA) calculation for
the transverse suceptibility with the solution
of the Boch's equations\cite{AJL+87}.
Using Eq.~(\ref{correlation}) into Eq.~(\ref{slichter})
or the RPA result,
we evaluate the zero temperature decoherence rates
due to the pseudo-spins as
\begin{eqnarray}
T_{1}^{-1}=T_{2}^{-1} & \approxeq  &  8 \pi^{-1} \gamma _{N}\left(\hslash \tilde{d} \gamma _{s}^{z}\right)^{2}B_{z}^{3} J_{\perp }^{-4}.\label{eq:t2}
\end{eqnarray}
The unusual dependence of the relaxation
time with the magnetic field, scaling like $B_{z}^{3}$, can be used
to assert Eq.~(\ref{eq:hf-1}-\ref{eq:hype2}). Finally, an applied  transverse
electric field ($E_{x,y}\neq 0$) can be used to open a gap in the
pseudo-spin spectrum.
This further isolates the qubit by quenching the pseudo-spins magnetic moments,
and therefore, even smaller values of $T_{1,2}^{-1}$ can be achieved.

In general, nuclear spins interacting with a 
gapless spin chain
would have super-ohmic dissipation.
However, the hyperfine Hamiltonian, Eq.~(\ref{eq:hype2}), that we derive
depends exclusively on the $z$ component of the pseudo-spins.
This restricted dipolar interaction implies an ohmic dissipation.
We emphasize that this is somewhat unique feature of pseudo-spins.
If Eq.~(\ref{eq:hype2}) would have
flip-flop terms, then the transverse correlations of the
spseudo-spins would imply a super-ohmic behavior.

\subsection{Construction of quantum gates}

\begin{figure}[tbh]
\includegraphics[  width=0.9\columnwidth]{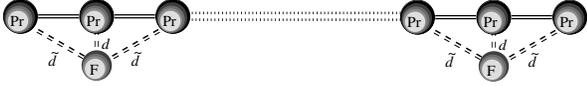}
\caption{\label{cap:interaction.}Interaction between two nuclear spins of
$\textrm{F}$ ions is mediated by the pseudo-spin chain of $\textrm{Pr}$.
$d$ and $\tilde{d}$ are the strength of the hyperfine coupling
defined in Eq.~(\ref{eq:hf-1}).}
\end{figure}

Now that we have studied the single qubit problem, we turn our
attention to the qubit-qubit interaction. We focus in the regime described by
Eq.~(\ref{eq:hype2}) because it is the most favorable for QC.
Consider a second $\textrm{F}$ atom along the
chain as shown in Fig.~(\ref{cap:interaction.}).
By integrating out the $\sigma $-spins we obtain a retarded interaction
between the two nuclei. This is  very similar to the RKKY interaction,
but mediated by the pseudo-spins \cite{RKKY}.

Exactally as in the RKKY problem, the $\textrm{F}$ nuclear spins have a
much slower dynamics than the pseudo-spins
($\gamma_{N}B_z \ll  J_\perp / \hslash$).
Therefore, it is reasonable to consider an instantaneous approximation to 
the interaction.
At zero temperature, we use 
Eq.~(\ref{correlation}) to calculate its form. 

For the RKKY,  finite temperature corrections are usually irrelevant
because the 
Fermi energy is much larger than the temperatures under consideration.
However, in the pseudo-spin
chain we are assuming temperatures only one order of magnitude smaller
than $J_{\perp}$.
We can easily re-write the zero temperature correlation function,
Eq.~(\ref{correlation}), in its finite temperature form by using the 
conformal invariance of the $\textrm{XY}$ model \cite{SEIA94}.
The final result is the effective
interaction between to qubits
\begin{eqnarray}
H_{\textrm{eff}} & \approxeq  & f_{zz}I_{1}^{z}I_{2}^{z}+f_{yz}\left(I_{1}^{y}I_{2}^{z}+I_{1}^{z}I_{2}^{y}\right)+f_{yy}I_{1}^{y}I_{2}^{y},\label{eq:Heff}
\end{eqnarray}
where we have defined the nuclear exchange couplings
%\begin{subequations}
\begin{eqnarray*}
f_{zz} & = & d^{2}G\left(\Delta x\right)-d\tilde{d}\left[G\left(\Delta x-1\right)+G\left(\Delta x+1\right)\right]\nonumber \\
 & + & \tilde{d}^{2}\left[2G\left(\Delta x\right)+G\left(\Delta x+2\right)+G\left(\Delta x-2\right)\right],\\
f_{yz} & = & \sqrt{2}\left\{ d\tilde{d}\left[G\left(\Delta x-1\right)-G\left(\Delta x+1\right)\right]\right.\nonumber \\ 
 & + & \left.\tilde{d}^{2}\left[G\left(\Delta x+2\right)-G\left(\Delta x-2\right)\right]\right\} ,\\
f_{yy} & = & 2\tilde{d}^{2}\left[2G\left(\Delta x\right)-G\left(\Delta x+2\right)-G\left(\Delta x-2\right)\right].
%\label{exchange}
\end{eqnarray*}
%\label{exchange}
%\end{subequations}
$G(\Delta x)$ is the finite temperature pseudo-spin propagator given by
\begin{eqnarray}
 G\left(\Delta x\right) & \approxeq & \frac{1-\cos \left(2p_{F}\Delta
    x\right)}{2\pi ^{2}v^{3}\beta ^{2}}\left[\sinh \left(\frac{\Delta
      x}{v\beta }\right)\right]^{-2} \, ,
\label{propagator}
\end{eqnarray}
where $\Delta x$ is the distance between qubits in units of lattice spacing
$a_{0}$ and $\beta =1/\left(k_{B}T\right)$.
For distances smaller than the thermal coherence length, $\xi _{T}=v\beta $,
the interaction decays as a power law, $G\left(\Delta x\right)\approxeq 
(1-\cos \left[2p_{F}\Delta x\right])/(2\pi ^{2}v\Delta x^{2})$,
leading to long range interaction between qubits. 
It is also interesting to consider the consequences of applying transverse
electric fields. 
Since the pseudo-spin propagator acquires a gap, there is an additional exponential
decay in Eq.~(\ref{eq:Heff}) which is a function of $E_{(x,y)}$.
Thus, we can use transverse fields to switch on and off the interaction between qubits. 

Equation~(\ref{eq:Heff}) is a two-qubit gate. In conjunction with the
possibility to perform arbitrary rotations, it generates a complete set of
quantum gates \cite{SL95}. 
The inverse of the gate operation time is
given by $T_{G}^{-1}\left(\Delta x\right)=\hslash ^{-1}\min \left(\left|f_{zz}\right|,\left|f_{yz}\right|\right)$.
In order to compare $T_{G}$ with $T_{1,2}$ we consider
a particular case. Take $B_{z}\approx 2\textrm{T}$ and a temperature
$T=0.1\textrm{K}$,
so that the pseudo-spin chain is partially polarized.
Low temperature corrections
to Eq.~(\ref{eq:t2}) are very small, and we use it as an upper
bound estimate to the decoherence times,
%in the case
%of \emph{perfect decoupling} and strong $E^{z}$:
$T_{1,2}^{-1}\sim 10^{-2}\textrm{s}^{\textrm{-1}}$.
These values are much smaller than the rates in  $\textrm{PrBr}_{\textrm{3}}$
% the parent compound
due to three facts: the absence of quadrupolar effects, the reduction of pseudo-spin fluctuation
in  $\tau _{0,1}$ and the assumption of decoupling.
Two qubits separated by $13\textrm{Å}$
have $T_{G}\left(3\right)\sim 10^{-1}\textrm{s}$, thus leading to a quantum 
gate at the edge of the error correction threshold of $10^{-4}$
\cite{LMKV02,errorcode}.

Another important aspect of Eqs.~(\ref{eq:fullhamispin}) and (\ref{eq:hf-1})
is that several different gates can be constructed as a function of
the magnetic field $B_{z}$, the resonance frequencies
$\omega _{\Delta ,\bar{\Delta }}$
and the electric fields $E_{x,z}$.
For instance,
the pseudo-spin propagator,
Eq.~(\ref{propagator}), has an oscillatory behavior with $B_{z}$.
This can be used to change the relative
strength of $f_{\textrm{ij}}$ in Eq.~(\ref{eq:Heff}).
In order to make this point clear, we now pause and consider a concrete
example.

One of the most simple quantum circuits is the one that creates
entangle pairs of qubits (Bell's states). From the quantum-logic perspective,
this is accomplished by the use of a Hadamard gate follow by a CNOT
gate \cite{NCh}. Since the production of entangle pairs
is fundamental to perform quantum computation and quantum comunication,
this straightforward circuit is a conerstone in any design.
The key element here is the CNOT gate.It is a two qubit gate and,
consequently its implementation depends upon the avaiable interaction.
In liquid state NMR the strongest component in the
Hamiltonian that a pair of qubits is subjected is \cite{LMKV02}
\begin{eqnarray}
H_{z} & \cong  & JI_{1}^{z}I_{2}^{z}.\label{eq:gate1}
\end{eqnarray}
This Hamiltonian can also be approximated by Eq.~(\ref{eq:Heff}).
For the sake
of argument, let us assume two F atoms separated by four lattice sites
($\sim 18\textrm{Å}$). In addition, let us consider the external conditions
that we considered before: a large $E^{z}$ to freeze the pseudo-spin dynamics
in $\tau _{0,1}$ and $T=0.1K$.

\begin{figure}[htb]
\includegraphics[
  width=1\columnwidth,
  keepaspectratio]{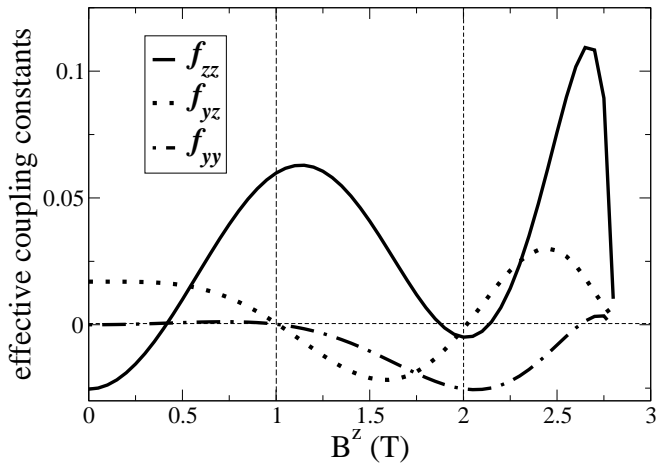}
\caption{\label{cap:the-coupling-constants}the coupling constants
$f_{zz}$,$f_{yz}$
and $f_{yy}$ as a function of the magnetic field $B^{z}$ in units
of $\tilde{d}^{2}$ for two F atoms separated by four lattice spacing.}
\end{figure}

From Eq.~(\ref{eq:Heff}) and the definition of the pseudo-spin
propagator we can plot Fig.~(\ref{cap:the-coupling-constants}),
where we can see that for $B^{z}\cong 1.01T$ the effective coupling
constants are
\begin{eqnarray*}
\frac{f_{zz}}{\tilde{d}^{2}} & \cong  & 0.06,\\
\frac{f_{yz}}{\tilde{d}^{2}} & \cong  & 0.001,\\
\frac{f_{yy}}{\tilde{d}^{2}} & \cong  & 0.001.
\end{eqnarray*}

Thus, as in liquid NMR, the strongest component in the interaction
is given by Eq.~(\ref{eq:gate1}). In order to produce a a CNOT gate with
this Hamiltonian in an NMR setup\cite{LMKV02}, one first apply
a radio frequency
pulse to rotate $I_{2}$ about $\hat{x}$ ( $+\hat{z}$ goes to $-\hat{y}$).
Then the spin system evolves with Eq.~(\ref{eq:gate1}) for a time
$t=\pi \hslash /4J$.
Then, a second pulse is sent to rotate $I_{2}$ by $90^{\circ }$ about the
$-\hat{y}$ axis. Finally, an additional phase shift on both spins
is used to obtain the CNOT gate. If we add the initial Hadamard gate,
it is necessary to use a total of five radio frequency pulses
(one-qubit gates) and
the time evolution of the Hamiltonian Eq.~(\ref{eq:gate1}) .

Let us analyze another possibility. Consider the same condictions as before,
but with an external magnetic field $B^{z}=2.02T$. In this case, the effective
coupling constants are

\begin{eqnarray*}
\frac{f_{zz}}{\tilde{d}^{2}} & \cong  & -0.005,\\
\frac{f_{yz}}{\tilde{d}^{2}} & \cong  & \phantom +0.0003,\\
\frac{f_{yy}}{\tilde{d}^{2}} & \cong  & -0.025.
\end{eqnarray*}

In contrast with the previous case, the strongest part of the interaction is

\begin{eqnarray}
H_{y} & \cong  & f_{yy}I_{1}^{y}I_{2}^{y}.\label{eq:gate2}
\end{eqnarray}

If we allow a free evolution of the system by Eq.~\ref{eq:gate2} for
a time $t=\pi \hslash /2f_{yy}$, the unitary transformation that
is implemented is:

\begin{eqnarray*}
R & = & \left[\begin{array}{cccc}
 1 & 0 & 0 & -i\\
 0 & 1 & i & 0\\
 0 & i & 1 & 0\\
 -i & 0 & 0 & 1\end{array}
\right]
\end{eqnarray*}

Acting on the computational basis with this rotation we automatically
generate the entangled states

\begin{eqnarray*}
\left|\beta _{1}\right\rangle  & = & \frac{\sqrt{2}}{2}\left(\left|00\right\rangle -i\left|11\right\rangle \right),\\
\left|\beta _{2}\right\rangle  & = & \frac{\sqrt{2}}{2}\left(\left|01\right\rangle +i\left|10\right\rangle \right),\\
\left|\beta _{3}\right\rangle  & = & \frac{\sqrt{2}}{2}\left(\left|01\right\rangle -i\left|10\right\rangle \right),\\
\left|\beta _{4}\right\rangle  & = & \frac{\sqrt{2}}{2}\left(\left|00\right\rangle +i\left|11\right\rangle \right).
\end{eqnarray*}

Hence, one can fine tune the experimental setup to obtain a desired 
quantum circuit using less resources.
In the above example, the simple tuning of the
magnetic field replace the one qubit gates on the previous setting.
However, this is just one of many possible ways to control
the interaction Hamiltonian. A more subtle (and potentially more interesting
way) is related to the frequency $\omega _{\Delta }$ and
$\omega _{\bar{\Delta }}$.
In presence of a gradient magnetic field they have a
site index
($\hslash \omega _{\Delta }\cong \sqrt{\left[\hslash \gamma _{s}^{z}B^{z}\left(\vec{x}\right)\right]^{2}+\Delta ^{2}}$).
Thus, one could act in the magnetic environment of
each individual qubit.

\section{Discussion and Conclusions} 

Until this point we discussed how single qubits can be constructed and how
a pair of qubits can interact.
We now discuss how to use these building blocks in a QC.

The natural geometry is to consider a magnetic field gradient applied
along the chain direction.
Nuclei in the same equipotential line belong to different copies of the QC,
and we assume that they can be periodically arranged (see below).

Initialization is a very hard problem in QCs based on nuclear spin qubits.
However, there are some possible solutions already available in the
literature \cite{AGTS+03,LJSUVV99}.
At first sight one could imagine that the initialization could
be done by optical pumping (Pound-Overhauser effect) with the pseudo-spins,
as it is done in  $MnF_{2}$ with electronic spin. Unfortunately,
the same property that gives a lower decoherence rate
than in other gapless magnetic systems hinders this option.
Since there
is no flip-flop term ($S^{+}I^{-}$) in the hyperfine Hamiltonian,
one cannot use the pseudo-spins to pump the nuclear spins.
There are two other possible {}``hardware'' solutions that can be
used to solve the initialization problem.
A diluted set of magnetic
impurities can be used to refrigerate the qubits. The general idea
is to add a small amount of an ion with a large magnetic moment (such as
$Gd$ replacing some $Pr$) to the sample. This set of impurities
can be used to pump energy out of the nuclear systems and after
some polarization is achieved a sufficiently large magnetic field
would {}``freeze'' the impurities. There are two setbacks
in this approach. Firstly, the $Gd$ ion would {}``break'' the pseudo-spin
chains and the F ions in each side might not interact. Secondly, virtual
flips of the $Gd$ spin could introduce an additional decoherence
channel. The second {}``hardware'' solution is based on the fact
that the crystals can be grown on a semiconductor substrate. By
exciting the electron gas in the semiconductor, it is possible to
use {}``cross-polarization-coherent transfer techniques''. The latter
is the solution found in ref.~\cite{AGTS+03}  to the initialization
procedure in a QC based on 1-d organic molecules. Finally, if only partial
polarization is obtained by one of the {}``hardware'' methods cited above,
the Schulman-Vazirani
procedure\cite{LJSUVV99} can be used as a {}``software''
method to initialize the state.

The final element in a QC design is the read-out mechanism.
All QCs based on NMR of impurities have the common problem of
low signal due to the small density of qubit copies.
However, nuclear polarization can increase considerably the NMR sensitivity.
In this case, the read-out  of a qubit with only $10^{12}$ copies
is possible with current NMR technology \cite{AGTS+03}.

There are some relevant experimental questions that are open
and can foster new theoretical work.
In the first place, the simplest way to produce crystals of a salt such as
$\textrm{PrBr}_{3-x}\textrm{F}_{x}$ is through dehydration of a liquid solution \cite{CMU69}.
This straightforward process creates samples with
the F ions in random positions.
Although this is sufficient to infer our results for a single qubit,
further developments
in ionic crystal growth should be accomplished
before the full range of possibilities
that we discuss can be experimentally studied.
One possible research avenue is a molecular-beam
epitaxial growth (MBEG).
MBEG is a well established
technique in semi-conductors and metals. Although from a historical
perspective the growth of ionic crystals is an old field, 
the technology is much less mature.
Nevertheless, it shows unique characteristics that are worth
exploring \cite{MBEG1,YF90, MBEG2}.
The most interesting feature
is that the incoming molecule has a very weak bound with the surface terrace
and strong bounding to the ledge.
This can be simply understood in electrostatic terms, and as a consequence,
leads to a large surface diffusion until the molecule reaches the
ledge. We speculate that this fact can be used to obtain a higher
degree of control in the impurity placement than in any other kind
of material. Another interesting characteristic is that large lattice
misfits are also allowed in the growth of layers. Thus, it is
natural to propose experiments with a crystal composed of a super-lattice
of $PrBr_{3}$ and layers of $PrBr_{2}F$. This setup
is feasible with the current technology and many of our results for the
qubit-qubit interaction can be experimentally tested. Another possibility
is a super-lattice of $PrBr_{3}$ with layers of $PrF_{3}$, however
the large lattice misfit will probably prevent the layer growth\cite{MBEG2}.
A final remark
is that ionic crystals grow well on semi-conductors surfaces. This
has two main consequences: 1) the semiconducting substrate can be integrated 
in other quantum
computer schemes (similar to Si/P proposals) and with current electronics,
2) a semi-conductor substrate can be used to initialize the quantum
computer by optical pumping as we argued above.

There isa another issue that is common to all solid state NMR designs:
it is unlikely that \emph{perfect decoupling} can be achieved.
Therefore, the experimental value of
$T_{2}$ is potentially smaller than the prediction of Eq.~(\ref{eq:t2}).
Although we are probably overestimating $T_{2}$, we are also underestimating
the gate time $T_{G}$. In order to derive Eq.~(\ref{eq:hf-1}),
we assumed an specific form to the hyperfine interaction. Following
the experimental results in $PrBr_{3}$ and $PrCl_{3}$, we assumed
that the dipolar part is the most relevant component in the hyperfine
Hamiltonian.
This conclusion arises from the hypothesis that the chemical bound is
truly ionic.
In general there are some covalent components to the bound and this
leads to a much stronger interaction with the electronic moments of
adjacent ions. For instance, this is preciselly what happens in $MnF_{2}$
\cite{AABB70,MF2}.
Whereas our hypothesis is based on the experimental facts in
$\textrm{PrBr}_{\textrm{3}}$
%the parent compound
\cite{SSRLA92,SSRLA+91,MDRLA83},
a thorough experimental study should be done to assert the hyperfine
Hamiltonian.

In summary, we showed how a non-Kramers ionic crystal has unique properties
that can be exploited in a solid state NMR QC.
We propose that chemical substitutions in such system can be used to encode
quantum information 
and, at the same time, break the spatial symmetries.
This \emph{controllable} symmetry-breaking can be used to act locally
in the magnetic environment of the qubit, thus, having important consequences
to decoherence and the construction of quantum gates.  
We based our discussion in a well known family of materials.
However, the general principle that we put 
forward can be applied in a much broader context.
In $\textrm{PrBr}_{\textrm{3-x}}\textrm{F}_\textrm{x}$, we
showed that a QC based on our ideas is scalable,
the decoherence rates are low, the interactions between qubits can be long ranged
and the qubits can be individually accessed with moderate magnetic field gradients.

\begin{acknowledgments}
We would like to thank I.~Affleck, C.~Chamon, N.~Curro, C.~Hammel,
D.~Loss, E.~Mucciolo, M.~Silva Neto, R. de Sousa, and D.~Taylor
for illuminating discussions.
%and the referees for the critical review of our manuscript and suggestions.

\end{acknowledgments}

\end{document}